# Simultaneous light emission and detection of InGaN/GaN multiple quantum well diodes for in-plane visible light communication on a chip

Yongjin Wang[1], Yin Xu[1], Yongchao Yang[1], Xumin Gao[1], Bingcheng Zhu[1], Wei Cai[1], Jialei Yuan[1], Rong Zhang[2] and Hongbo Zhu[1]

[1]Grünberg Research Centre, Nanjing University of Posts and Telecommunications, China

[2]School of Electronics and Computer Science, University of Southampton, UK

***Abstract***

This paper presents the design, fabrication, and experimental characterization of monolithically integrated p–n junction InGaN/GaN multiple quantum well diodes (MQWDs) and suspended waveguides. Suspended MQWDs can be used as transmitters and receivers simultaneously, and suspended waveguides are used for light coupling to create an in-plane visible light communication system. Compared to the waveguide with separation trench, the calculated total light efficiency is increased from 18% to 22% for the continuous waveguide. The MQWDs are characterized by their typical current-voltage performance, and the pulse excitation measurements confirm that the InGaN/GaN MQWDs can achieve the light emission and photodetection at the same time. The photocurrent measurements indicate that the photocurrent is modulated by a bias voltage and that the photons are being supplied from another transmitter. An experimental demonstration is presented showing that the proposed device works well for in-plane full-duplex communication using visible light.

## *Introduction*

For radio frequency (RF) communications, it was once considered impossible to receive and transmit on the same channel because of self-interference [1]. However, some recent works indicate that self-interference cancellation can be used to realize full-duplex RF communications [2-5], which can double the throughput of the entire communication system. Extracting a useful signal from the received distorted signal is challenging because the power of the unwanted self-interference can be billions of times stronger (100 dB+) than that of the desired signal. Therefore, highly accurate analog and digital circuits must work cooperatively to remove the self-interference. Such sensitive and expensive structures may not be practical for cost-sensitive situations, e.g., in portable devices. Are there any simpler ways to achieve full-duplex communications? Visible light communications (VLC) might be the answer.

In a VLC system, the emitter emits modulated light to deliver signals via visible light, and the detector senses the light through photon–electron conversion and thus extracts the signals [6-10]. Normally, the detector that absorbs the light cannot also emit light. However, a p–n junction InGaN/GaN multiple quantum well diode (MQWD) can perform both light emission and photodetection [11,12], enabling the realization of full-duplex communication on the same channel using visible light.

In-plane VLC is achieved through light propagation between the emitter and the detector rather than across short electrical wires [13-17]. In our previous work, the modulated light from the emitter was guided by suspended waveguides and transported to the detector. An isolation trench located in the waveguide was used to separate the two p-GaN regions in the in-plane VLC system, leading to an abrupt change in the light propagation along the waveguides [18-20]. Here, we report the realization of an in-plane VLC system consisting of two InGaN/GaN MQWDs and suspended waveguides on a single chip, fabricated on a 2-inch GaN-on-silicon platform using a wafer-level technique. Both the p-GaN and InGaN/GaN multiple quantum well (MQW) layers on top of the waveguides are removed to generate a continuous waveguide architecture. The waveguide height is thus reduced, and the fabrication process is simpler, especially for further nanoscale photonic integration. Simulations and experiments are conducted to characterize the proposed photonic integrated circuit.

***Methods***

The in-plane VLC system is implemented on a 2-inch GaN-on-silicon wafer [21, 22], which is thinned to 200-μm thickness for further silicon removal by chemical mechanical polishing. The device layers consist of a ~220-nm-thick p-GaN layer, 250-nm-thick InGaN/GaN MQWs, a ~3.2-μm-thick n-GaN layer, a ~400-nm-thick undoped GaN layer, a ~900-nm-thick Al(Ga)N buffer layer, and a silicon substrate. Because the InGaN/GaN MQWDs serve as LED and photodiode simultaneously, the fabrication technology for creating the integrated optoelectronic device is identical to that for an InGaN/GaN MQWs-based LED and photodiode. The top layer is defined by photolithography and etched down to n-GaN to form an isolation mesa by inductively coupled plasma reactive ion etching (ICP-RIE) with $Cl_2$ and $BCl_3$ hybrid gases. A thick p-contact is used to suppress the light emission from the top escape cone. Both p- and n-type contacts are then formed by Ni/Au (20 nm/180 nm) evaporation, followed by a rapid thermal annealing at 500 °C in an $N_2$ atmosphere. Next, the waveguides, in which both the top p-GaN layer and the InGaN/GaN MQWs layers are removed to form the isolation for the two InGaN/GaN MQWDs, are patterned and etched by ICP-RIE. The top device structures are protected by thick photoresist and the silicon substrate is patterned by back-side alignment photolithography. Deep reactive ion etching is conducted to remove the silicon substrate, and back wafer etching of the suspended membrane is performed to obtain a membrane-type in-plane VLC system.

Figure 1(a) shows a scanning electron microscope (SEM) image of the fabricated in-plane VLC system. Devices A and B, which are used as transmitters and receivers simultaneously, are suspended InGaN/GaN MQWDs. The 70-μm-diameter p-electrodes are fabricated on the isolation mesa, which has a 10-μm-wide gap to the n-electrode. The light coupling between device A and device B is realized through the in-plane connection by three 80-μm-long, 2.47-μm high, and 10-μm-wide suspended waveguides. Compared with the previous waveguide structures, the waveguide height is reduced, leading to a decrease in the confined optical modes. Figure 1(b) illustrates a three-dimensional atomic force microscope (AFM) image of the suspended waveguides. The waveguide has a smooth top surface and the measured root mean square roughness is 1.16 nm, as obtained by an ICP-RIE process. The measured height of the isolation mesa is 525 nm.

***Results and Discussion***

The light propagation properties of the suspended waveguide are evaluated using the finite difference time domain method (FDTD). For simplicity, the refractive index of GaN used is 2.45. Because there are six escape cones for a planar emitter [23, 24], a point-like source with a wavelength of 450 nm is adopted for the simulation. Most of the emitted light is confined inside the suspended structure as a result of the large index contrast between GaN and air [25-29]. The light emitted from the emitter at the end of the waveguide couples into the suspended waveguide and laterally propagates along it. The detector at the other end of the waveguide senses the guided light and completes the photon–electron conversion. Figure 2 illustrates the FDTD simulation of the suspended waveguides. The light losses include the coupling losses between the emitter and waveguide, the coupling losses between the waveguide and detector, and the propagation loss inside the waveguide. The total light efficiency, defined by the ratio between the monitor power at the detector and the power at the emitter, is calculated to be around 18% for the waveguide with a 10-μm-wide and 530-nm-thick separation trench. The proposed continuous waveguide has an improved light efficiency to 22%.

Four-terminal measurements are applied to the in-plane VLC system by a combination of an Agilent B1500A semiconductor device analyzer and a Cascade PM5 probe station. Figure 2(a) shows the log-scaled current-voltage (I-V) plots for device B at the different injection current levels of device A. The I-V plot for device B demonstrates typical rectifying behavior with a turn-on voltage of ~2.5 V at the injection current of 0 mA for device A. As the injection current of device A is increased from 0.2 mA to 1 mA, the number of photons supplied by device A increases. These photons pass through the suspended waveguides and are absorbed by device B.

There the photons generate electron-hole pairs, the electrons flow to the n-contact and the holes move to the p-contact, leading to an improved photocurrent in the direction opposite to that of the current driven by the bias voltage. Hence, the measured current of device B is the sum of the driven current and the induced photocurrent. The currents are plotted on a log scale with their absolute values. Device B operates under detector mode below the turn-on voltage. The measured current, which is the induced photocurrent, increases with the injection current of device A because device A can then supply more photons for device B. The current increases with the bias voltage of device B, which can be attributed to the reduction in the barrier height of the p–n junction. Around the turn-on voltage, the current driven by the bias voltage is established, giving rise to a sharp change in the measured current. Device B operates in emitter mode above the turn-on voltage. Notably, a decrease in the measured current can be clearly observed with increases in the injection current of device A from 0.2 mA to 1 mA, indicating that the induced photocurrent is comparable to the driven current. The measured currents are further normalized with the current of device B at the forward current of 0 mA of device A to obtain the induced photocurrent. Figure 3(b) shows the normalized photocurrents. The induced photocurrent gradually increases with the injection current of device A when device B operates either under detector or emitter mode, indicating the nearly linear electron–photon conversion at device A and photon–electron conversion at device B. The measured photocurrent is over 1000 times stronger for device B under emitter mode than under detector mode. This means that the barrier manages to form a channel for photon–electron conversion when device B operates in emitter mode.

Similar results are observed when device A detects the light emitted from device B. These experimental results confirm that devices A and B can absorb the light guided by the suspended waveguides to complete the photon-electron conversion when the devices operate either under emitter mode or detector mode, leading to the realization of simultaneous light emission and photodetection.

When a pulse signal (50μs, 5V) is applied on the device B, an excited signal is observed at the device A with a bias voltage of 0 V, as illustrated in Figure 4 (a). When the pulse signal is finished, the excited signal gradually returns to the initial state. When the device A is applied with an injection current of 400 μA, the device A is turn-on and emits light. Figure 4(b) shows the measured excited signal. The signal amplitude is increased and the decay time is decreased. These results indicate that the InGaN/GaN MQWDs can achieve the light emission and photodetection at the same time.

To experimentally demonstrate that the proposed device works for in-plane full-duplex communication using visible light, devices A and B are both directly driven by an Agilent 33522A arbitrary waveform generator to modulate the emitted light. Photon-electron conversion and electron-photon conversion occur at each communication port simultaneously, and the signals are characterized by an Agilent DSO9254A digital storage oscilloscope without amplification. Figure 5(a) shows the transmitted signals at device A and the received signals at device B when a 0 V bias voltage is applied to device B and device A is driven at 3.0 V to modulate the emitted light as a sine wave signal at 13 MHz. In this case, device B operates under detector mode. When the bias voltage is increased to 3.0 V, device B operates under emitter mode. Figure 5(b) illustrates that the amplitudes of the received signals are correspondingly enhanced. Figure 5(c) shows the amplitude of the received signal versus the bias voltage of device B. As a receiver, device B can detect the light either under emitter mode or under detector mode, and the received signals are stronger for device B under emitter mode than under detector mode, showing the same trend as in Figure 3(b). The in-plane visible light communication system is then characterized by a combination of a Keysight 81160A pulse function arbitrary generator and a Keysight N9010A EXA signal analyzer. As shown in Figure 5(d), experimental results confirm that the device can transmit a 500 MHz sinewave and receive a 300 MHz sinewave simultaneously, confirming that the proposed device can detect modulated light while also emitting modulated light. The offset voltages are 3.0 V and the peak-to-peak modulation voltages are 1 V, and 6 V, respectively. These results experimentally demonstrate that full-duplex communication can be achieved on the same channel for the development of an in-plane VLC system.

Full-duplex communication using the same frequency is experimentally demonstrated at 50 kHz because of the limitations of the probe station used. Under half-duplex mode, the receiver is at the bias voltage of 0 V and the emitter is directly driven by the signal generator to modulate the light at 50 kHz. The square wave signal used has a filling factor of 0.5. Figure 6(a) shows the signal traces at the emitter. When the in-plane VLC system operates in full-duplex mode, the communication port emits the modulated light and detects the light at the same time. At the other communication port, the emission is modulated at 50 kHz with a filling factor of 0.25. When both transmitted and received signals happens at the same time and at the same frequency, the superposition behavior occurs. As illustrated in Figure 6(b), both transmitted and received signals can be detected, indicating the in-plane full-duplex VLC system can operate at the same frequency.

***Conclusion***

In summary, we have presented the design, fabrication, and testing of a photonic integrated circuit composed of two InGaN/GaN MQWDs, as well as suspended waveguides for light coupling, to create an in-plane VLC system. First, FDTD simulation results suggest that the total light efficiency can be improved by the utilization of continuous waveguide structure. Second, the InGaN/GaN MQWDs are characterized by their typical I-V performance, and the pulse experiments confirm that the InGaN/GaN MQWDs can detect light either under emitter mode or detector mode to achieve light emission and photodetection simultaneously. The photocurrent is modulated by the bias voltage and the photons supplied from different emitters. The experimental results demonstrate that the proposed device works as designed for in-plane full-duplex communication using visible light. The techniques developed in our experiments will be important for creating an in-plane full-duplex VLC system.

***Acknowledgements***

This work is jointly supported by the National Natural Science Foundation of China (grant nos. 61322112 and 61531166004) and Research Projects (nos. 2014CB360507). The financial support of the EPSRC project (EP/N023862/1) is gratefully acknowledged.

**Illustration Captions**

Figure 1 (a) SEM image of the fabricated in-plane VLC system, (b) an AFM image of the suspended waveguide.

Figure 2 FDTD simulation of the suspended waveguide.

Figure 3 (a) Log-scaled I-V plots for device B with different device A injection current levels, (b) the normalized photocurrents of device B.

Figure 4 (a) Measured pulse signal at the bias voltage of 0 V, (b) Measured pulse signal at the bias voltage of 3.0 V.

Figure 5 (a) Measured signals for device B at the bias voltage of 0 V when device A is driven at 13 MHz, (b) the measured signals for device B at the bias voltage of 3 V when device A is driven at 13 MHz, (c) the amplitude of the received signal versus the bias voltage of device B, (d) the measured signals for the in-plane full-duplex VLC system when both devices A and B are driven at 500 MHz and 300 MHz, respectively.

Figure 6 (a) The in-plane VLC system operating in half-duplex mode, (b) the in-plane VLC system operating in full-duplex mode at the same frequency.

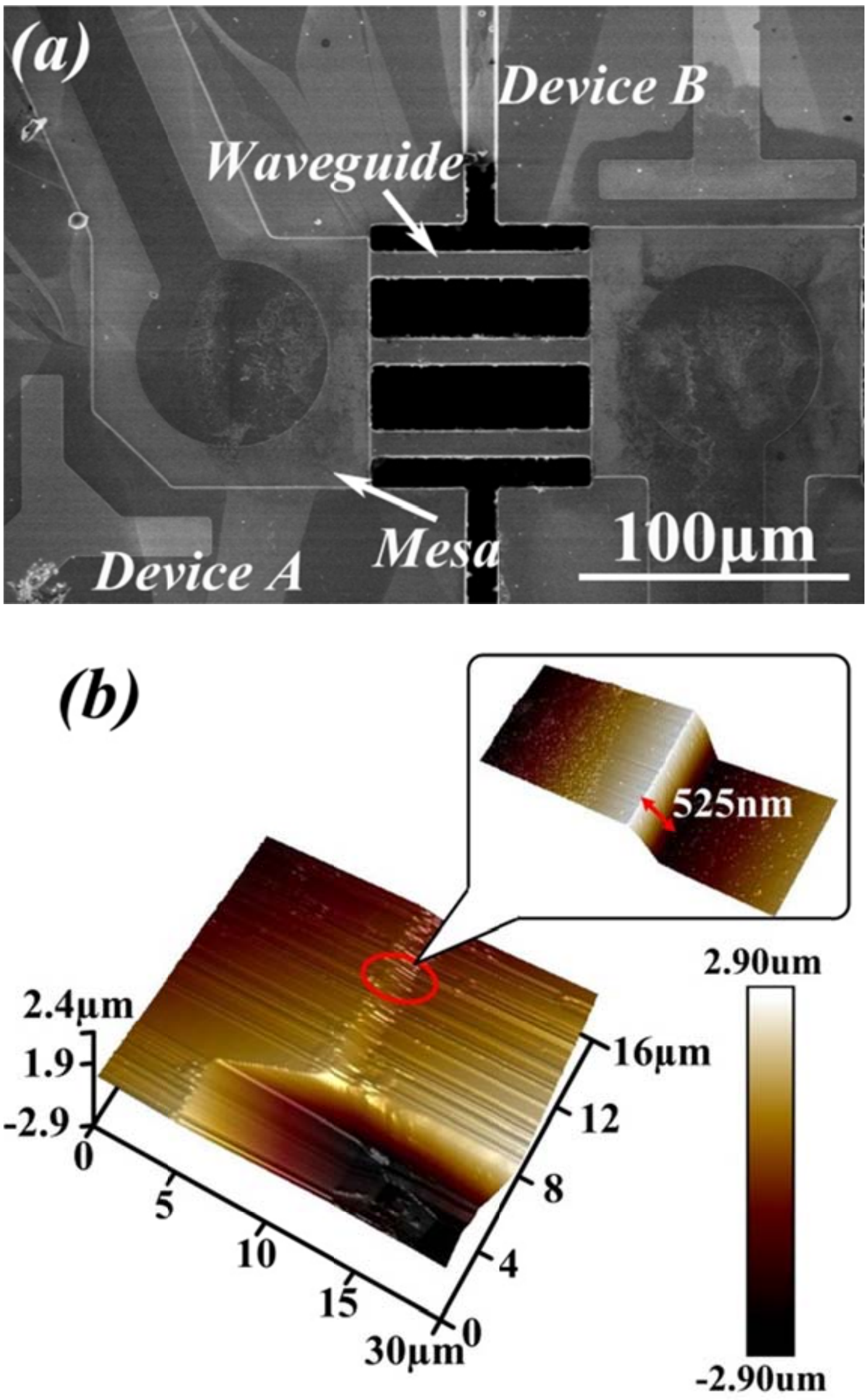

**Figure 1**

***Yongjin Wang et al.***

#

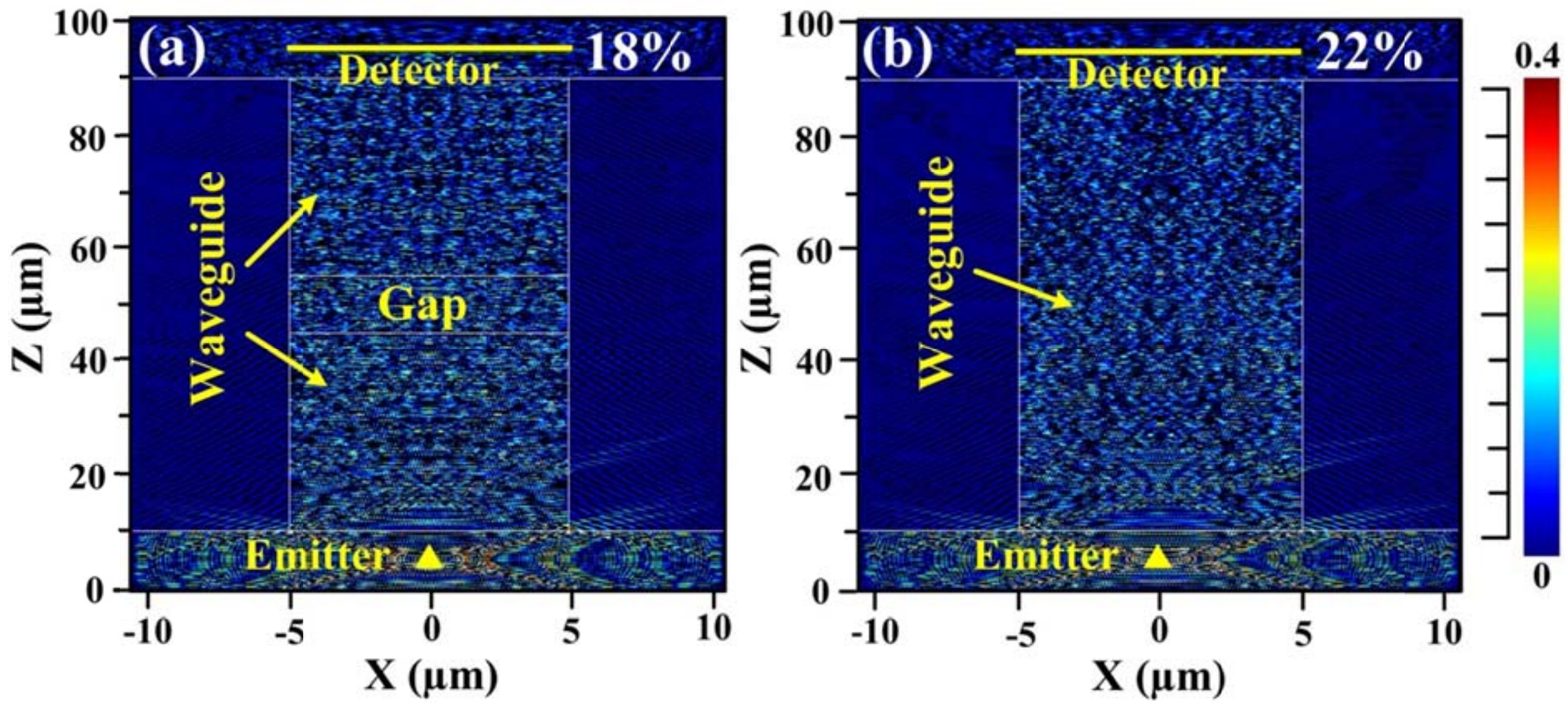

#

**Figure 2**
***Yongjin Wang et al.***

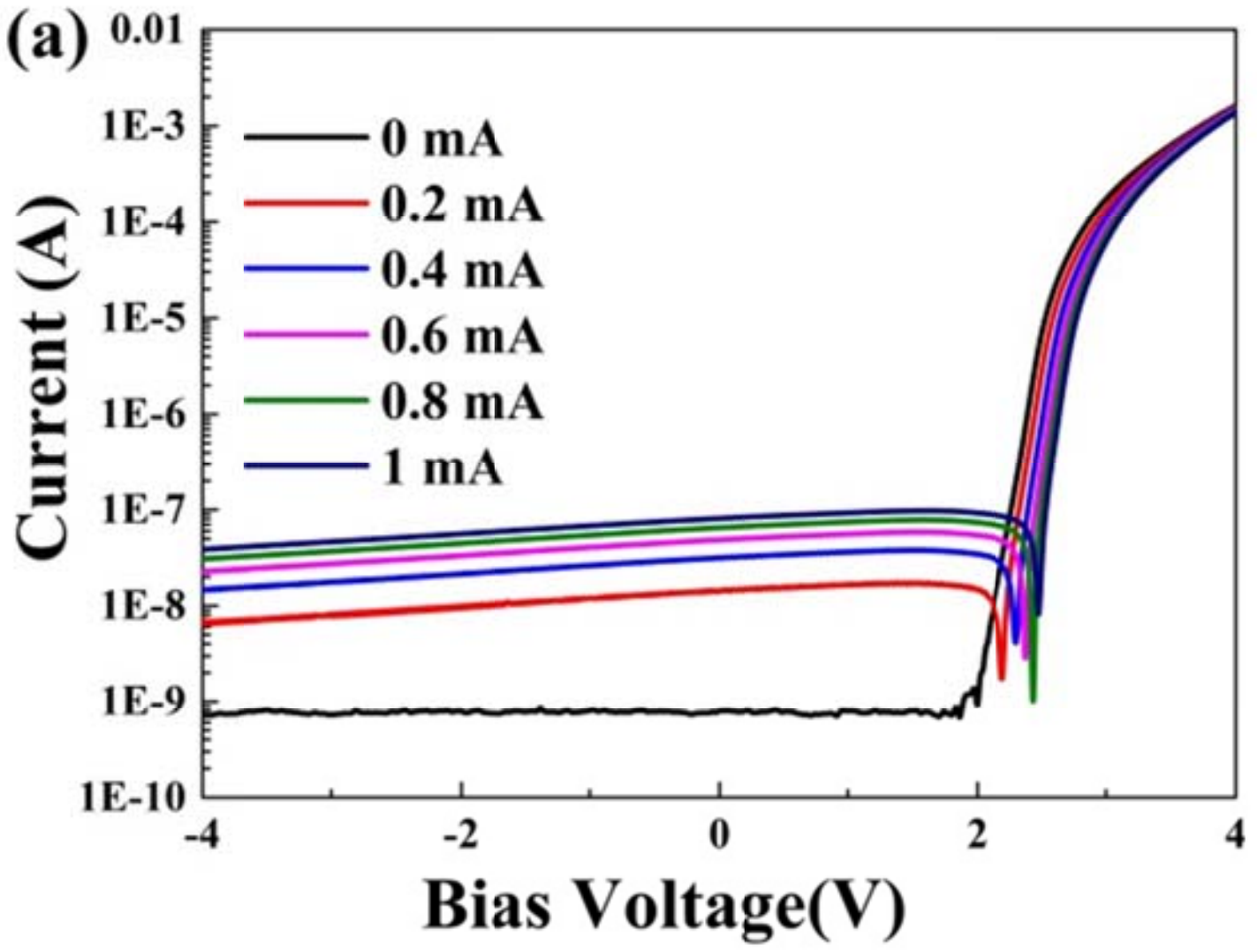

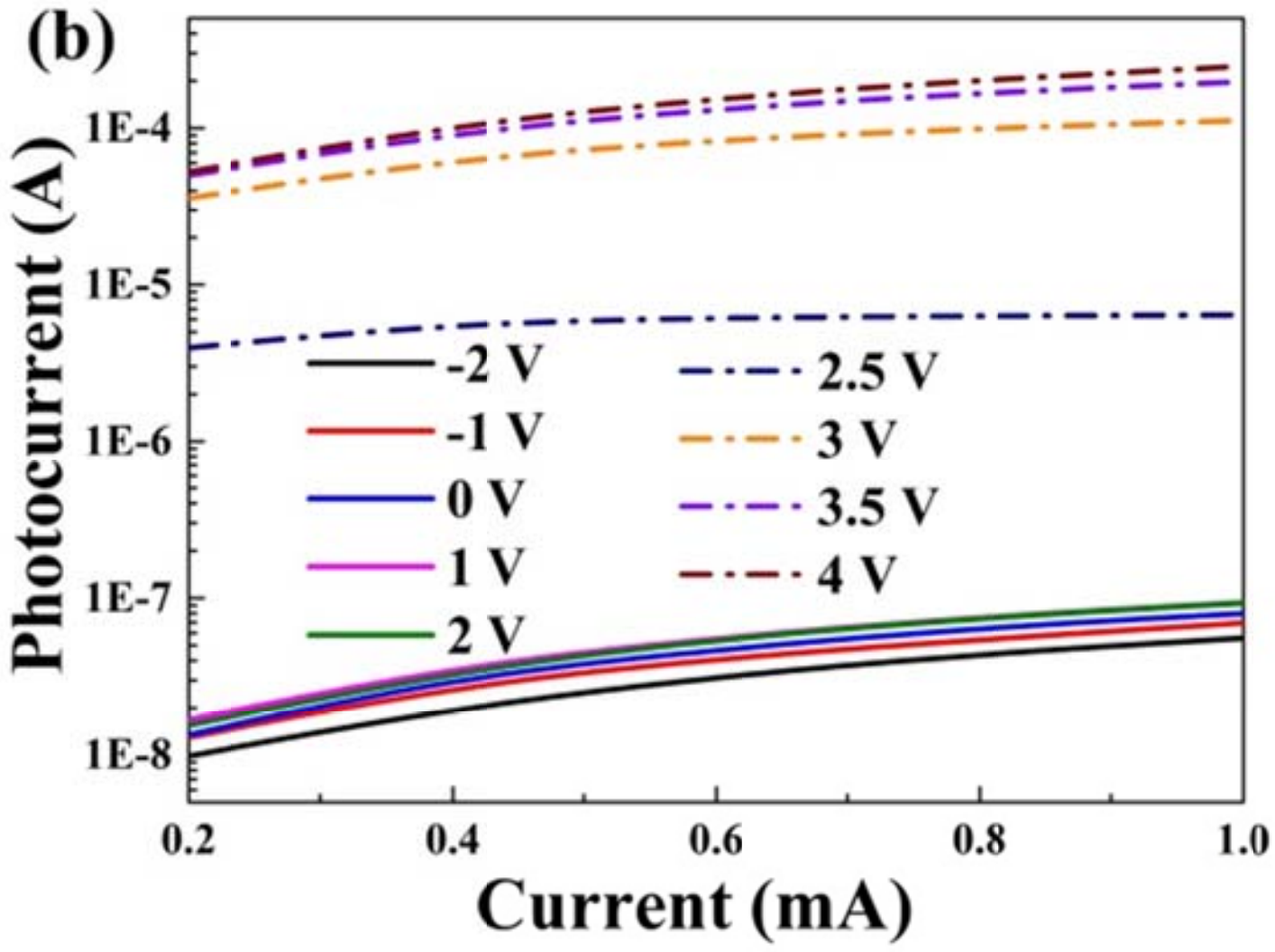

**Figure 3**

***Yongjin Wang et al.***

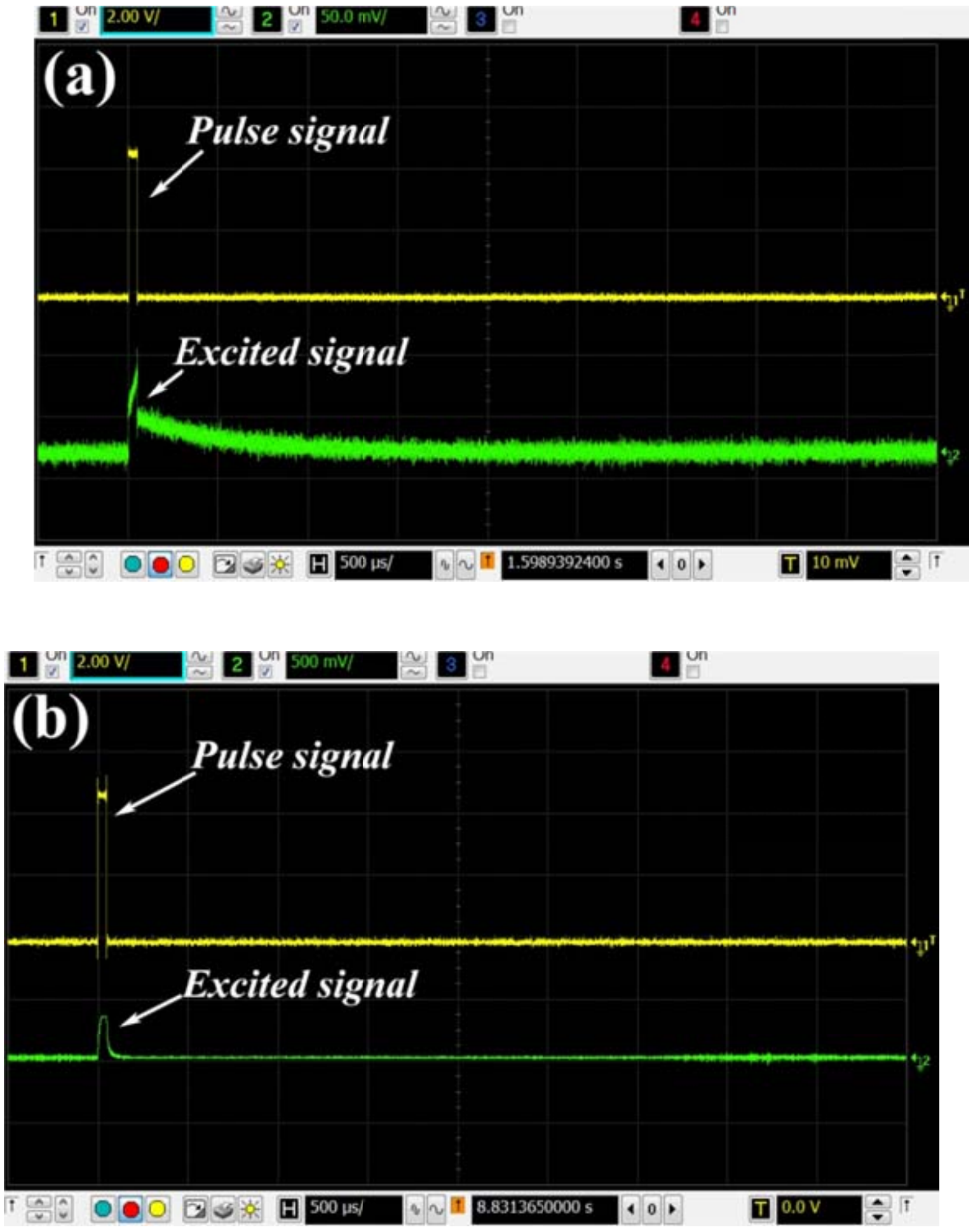

**Figure 4**
***Yongjin Wang et al.***

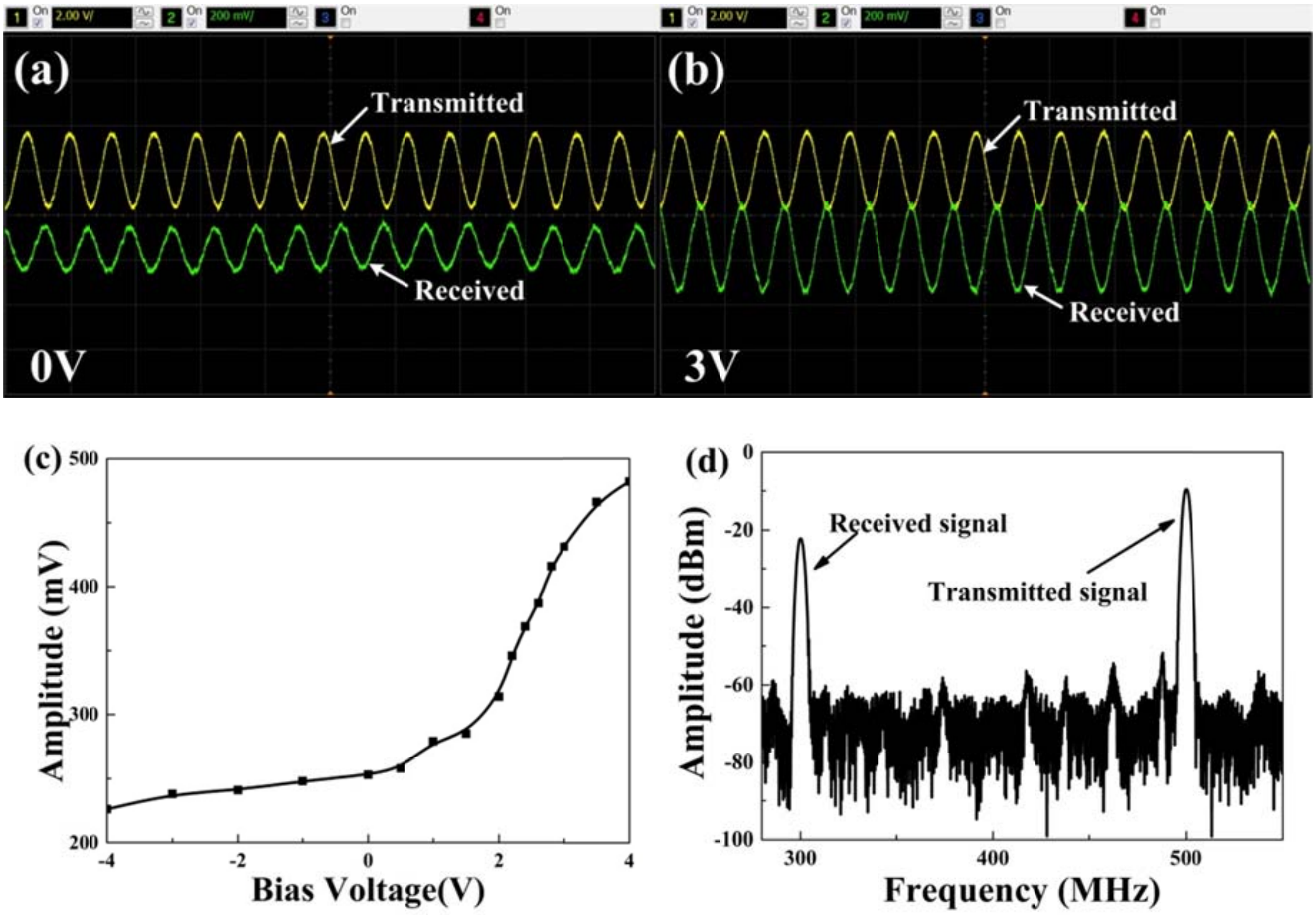

**Figure 5**

***Yongjin Wang et al.***

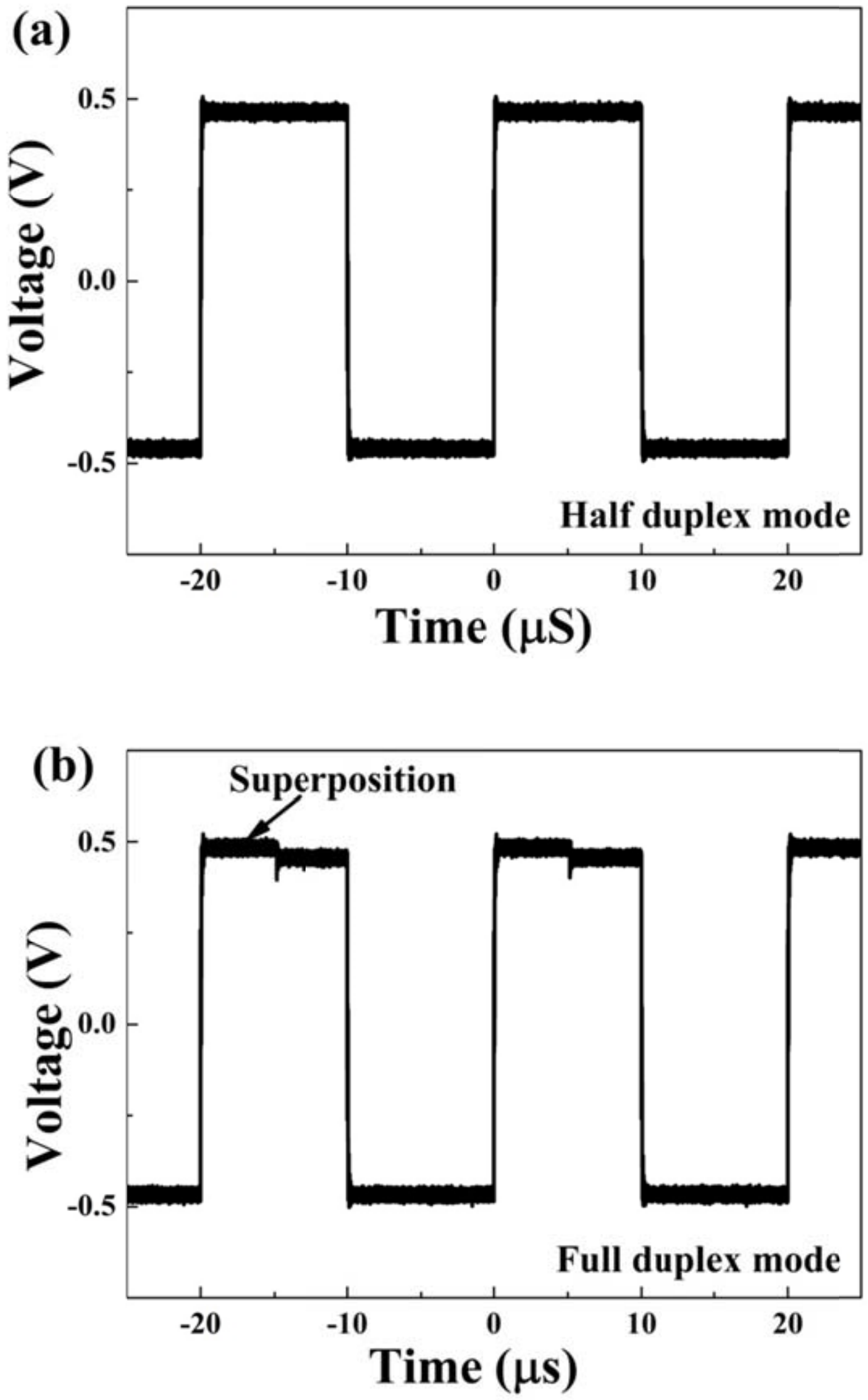

**Figure 6**

***Yongjin Wang et al.***